\documentstyle[aps,pre,epsfig,amsmath,amssymb,preprint]{revtex}

\newcommand{\be}{\begin{eqnarray}}
\newcommand{\ee}{\end{eqnarray}}

\begin{document}

\title{The Vortex Kinetics of Conserved and Non-conserved O(n) Models}
\author{Hai Qian and Gene F. Mazenko}
\address{James Franck Institute and Department of Physics, University of
Chicago, Chicago, Illinois 60637}
\date{today}

\maketitle

\begin{abstract}
We study the motion of vortices in the conserved and non-conserved
phase-ordering models. We give an analytical method for computing
the speed and position distribution functions for pairs of annihilating point
 vortices based on
heuristic scaling 
arguments. In
the non-conserved case this method produces a speed distribution function
consistent with previous analytic results. As two special
examples, we simulate the conserved and
non-conserved $O(2)$ model in two dimensional space numerically. The
numerical results for  the non-conserved case are consistent with the theoretical predictions. The
speed distribution of the vortices in the conserved case is measured for
the first time. Our theory produces a distribution function with the
correct large speed tail but does not accurately describe the numerical
data at small speeds. The position distribution functions for both models are
measured for the first time and we find good agreement with our
analytic results. We are also able to extend this method to models
with a scalar order parameter.
\end{abstract}




\section{Introduction}

The phase ordering dynamics of certain physical systems after a rapid
temperature quench
below its critical temperature is dominated by the annihilation of
topological defects of opposite charge \cite{Bray94}. In particular the
$n$-vector model 
with non-conserved order parameter (NCOP) Langevin dynamics, where the defects are vortices, 
has been studied in some detail \cite{Liu92,Mondello90,Mazenko97}. 

Mazenko \cite{Mazenko97,Mazenko97_2} carried out an investigation of the
distribution of defect velocities for non-conserved phase ordering systems. By
using an approximate ``Gaussian closure'' scheme, he was able to compute
the velocity distribution for vortices in the non-conserved $n$-vector
Langevin model for the case of point defects where $n=d$ dimensions.
 We \cite{Qian03} carried out numerical 
simulations for the $n=d=2$ non-conserved case and measured the vortices
speed distribution. The results are consistent
with Mazenko's theoretical predictions. In particular the power-law tail of the
distribution at large speeds which is robust is correctly predicted.  
The problem of the relative velocity as a function of separations for
annihilating pairs was treated in Ref. \cite{Mazenko97_2}, and the
velocity distribution  for strings for the non-conserved order parameter
case was  treated in Ref. \cite{Mazenko99}.

Bray \cite{Bray97} developed a heuristic scaling treatment of
the large speed tails based
on the disappearance of small defects (annihilating pairs or contracting
compact domains). This method treated only 
the power-law exponent of the distribution's large speed
tail. For the non-conserved $n$-vector case this simple argument gives
a result consistent with Mazenko's theory. However this
method is also able to produce the large speed tail exponents for the
conserved $n$-vector models, and conserved and non-conserved scalar
(n=1) models. 

We show here that Bray's arguments can be extended to give results
beyond the tail exponents. In models where point defects dominate the
dynamics,  one can compute the
defect speed distribution functions based on Bray's scaling assumption.
The same idea can be easily generalized to the scalar order parameter case.

In a very recent work Mazenko \cite{Mazenko03} has suggested how the
previous work in Ref. \cite{Mazenko97} can be extended to anisotropic
systems and the conserved order parameter (COP) case. He finds that the
average speed goes as $t^{-1}$ for the COP case with a scaling function
of the same formal form as for the NCOP case given by Eq. (\ref{NCOPv})
below. These results are not in agreement with the analytical or
numerical work presented below in this paper. The
Gaussian closure method developed in Ref. \cite{Mazenko03} does not
appear adequate for treating the COP case.

In the next section we will generalize Bray's argument for the point
defect case. We find simple analytic expressions for the speed and
separation  distribution functions. We recover the large speed tail
exponents obtained previously by Bray. For the non-conserved $n$-vector
model we obtain precisely the same results as found in
Ref. \cite{Mazenko97}. Then in Sec. III 
we present the numerical simulation results for non-conserved $n=d=2$
Langevin model. Next in Sec. IV, we present the simulations results for
the 
conserved $n=d=2$ Langevin model. In Sec. V,
we point out that the method developed in this paper can be used for those cases
where one has a scalar order parameter. 

\section{Theoretical Development}

Let us suppose that we have $N$ pairs of oppositely
charged vortices which are on their way to annihilation.  We
suppose that pair $i$ is separated by  distance
$r_{i}(t)$ with relative speed $v_{i}(t)=|\dot{r}_i(t)|$.  Consider
the associated phase-space distribution function:
\be
f(r,v,t)=\langle \sum_{i=1}^{N}
\delta (r -r_{i}(t))\delta (v -v_{i}(t))\rangle
~~~.
\ee
This quantity satisfies the equation of motion
\be
\frac{\partial }{\partial t}f(r,v,t)
=-\frac{\partial }{\partial r}
\langle \sum_{i=1}^{N}\dot{r}_{i}(t)
\delta (r -r_{i}(t))\delta (v -v_{i}(t))\rangle
-\frac{\partial }{\partial v}
\langle \sum_{i=1}^{N}\dot{v}_{i}(t)
\delta (r -r_{i}(t))\delta (v -v_{i}(t))\rangle
~~~.
\label{eq:2}
\ee
Our key kinematical assumption is that the relative velocity
is a known function of the separation:
\be
\dot{r}_{i}(t)=-v_{i}(t)=-u(r_{i}(t))
\ee
\be
\dot{v}_{i}(t)=u^{\prime}(r_{i}(t))\dot{r}_{i}(t)
=-u(r_{i}(t))u^{\prime}(r_{i}(t))
~~~.
\ee
We check these assumptions as we proceed. Eq.(\ref{eq:2}) then takes the form
\be
\frac{\partial }{\partial t}f(r,v,t)
=\frac{\partial }{\partial r}
\left(u(r)f(r,v,t)\right)
+\frac{\partial }{\partial v}
\left(u(r)u^{\prime}(r)f(r,v,t)\right)
~~~.
\label{eq:1}
\ee
where we have the normalization
\be
\int_{0}^{\infty}dr \int_{0}^{\infty}dv f(r,v,t)=N(t)
~~~.
\ee
Eq.(\ref{eq:1}) is one of our primary results.

Our assumptions are consistent with being in a regime
where the annihilating pairs are independent and we
can write
\be
f(r,v,t)=N(t)P(r,v,t)
\label{eq:7}
\ee
where $P(r,v,t)$ has the interpretation as the
probability that at time $t$ we have a pair separated
by a distance $r$ with relative speed $v$.  Inserting
Eq.(\ref{eq:7}) into Eq.(\ref{eq:1}) we find
that $P(r,v,t)$ satisfies
\be
\frac{\partial }{\partial t}P(r,v,t)
=\frac{\partial }{\partial r}
\left(u(r)P(r,v,t)\right)
+\frac{\partial }{\partial v}
\left(u(r)u^{\prime}(r)P(r,v,t)\right)
+\gamma P(r,v,t)
\label{eq:8}
\ee
where
\be
\gamma =-\frac{1}{N(t)}\dot{N}(t)
~~~.
\label{eq:9}
\ee
We will see that $\gamma$ (and $N(t)$) are determined
self-consistently by using scaling ideas.

We are interested in the reduced probability distributions
\be \label{pr}
P_r(r,t)=\int_{0}^{\infty}dv P(r,v,t)
~~~,
\ee
and
\be
P_v(v,t)=\int_{0}^{\infty}dr P(r,v,t)
\ee
with the overall normalization
\be
\int_{0}^{\infty}dr \int_{0}^{\infty}dv P(r,v,t)=1
~~~.
\ee

Our goal is to solve Eq.(\ref{eq:8}).  The first step is
to show that
\be
P(r,v,t)=P_r(r,t)\delta (v-u(r))
~~~.
\label{eq:4}
\ee
Inserting Eq.(\ref{eq:4}) into Eq.(\ref{eq:8}) we have
\be
\delta (v-u(r))\frac{\partial }{\partial t}P_r(r,t)
=\gamma \delta (v-u(r))P_r(r,t)
+\frac{\partial }{\partial r}
\left(u(r)\delta (v-u(r))P_r(r,t)\right)
\nonumber
\ee
\be
+\frac{\partial }{\partial v}
\left(u(r)u^{\prime}(r)\delta (v-u(r))P_r(r,t)\right)
\nonumber
\ee
\be
=\delta (v-u(r))\left(\gamma P_r(r,t)
+\frac{\partial }{\partial r}\left(u(r)P_r(r,t)\right)\right)
\nonumber
\ee
\be
+u(r)P_r(r,t)\left(\frac{\partial }{\partial r}
\delta (v-u(r))+\frac{\partial }{\partial v}
u^{\prime}(r)\delta (v-u(r))\right)
~~~.
\ee
Using the following identity
\be
\frac{\partial }{\partial r}\delta (v-u(r))
=\frac{\partial }{\partial r}\int
\frac{d\lambda}{2\pi}e^{i\lambda (v-u(r))}
\nonumber
\ee
\be
=\int\frac{d\lambda}{2\pi}(-i\lambda u^{\prime}(r)
e^{i\lambda (v-u(r))}
\nonumber
\ee
\be
=-u^{\prime}(r)\frac{\partial }{\partial v}\delta (v-u(r))
\ee
we find that Eq.(\ref{eq:4}) holds with $P_r(r,t)$ determined
by
\be
\frac{\partial }{\partial t}P_r(r,t)
=\gamma P_r(r,t)
+\frac{\partial }{\partial r}
\left(u(r)P_r(r,t)\right)
~~~.
\label{eq:6}
\ee
Imposing the normalization,
\be \label{eq:17}
\int_{0}^{\infty}dr  P_r(r,t)=1
~~~,
\ee
we find
on integrating Eq. (\ref{eq:6}) over $r$ that
\be
\gamma =\lim_{r\rightarrow 0} u(r)P_r(r,t)
~~~.
\label{eq:18}
\ee
Thus $\gamma$ and $N(t)$ are determined self-consistently
in terms of the solution to Eq. (\ref{eq:6}).

So far this has been for general $u(r)$, let us restrict
our subsequent work to the class of models where the relative velocity
is a power law in the separation distance:
\be \label{ur}
u=Ar^{-b}
\ee
where $A$ and $b$ are positive.
Next we assume that we can
find a scaling solution \cite{foot}
to Eq. (\ref{eq:6}) of the form
\be
P_r(r,t)=\frac{1}{L(t)}F(r/L(t))
\label{eq:19}
\ee
where the growth law $L(t)$ is to be determined.
Inserting this ansatz into Eq.(\ref{eq:6}) we obtain:
\be
-L^{b}\dot{L}\left(xF'+F\right)=Ax^{-b}F'-Abx^{-b-1}F
+L^{b+1}\gamma F
\label{eq:21}
\ee
where $x=r/L(t)$.  To achieve a scaling solution we
require
\be
L^{b}\dot{L}=AC
\label{eq:22}
\ee
and
\be
L^{b+1}\gamma(t) =AD
\ee
where $C$ and $D$ are time independent positive constants,
the factors of $A$ are included for convenience.
Eq. (\ref{eq:21}) then takes the form
\be
-C\left[xF'+F\right]=DF+x^{-b}F'-bx^{-b-1}F
~~~.
\ee
This has a solution
\be
F(x)=\frac{Bx^{b}}{(1+Cx^{1+b})^{\sigma}}
\label{eq:25}
\ee
where $B$ is an over all positive constant and
the exponent in the denominator is given
by
\be
\sigma =1+\alpha /z
\ee
where $z=1+b$ and $\alpha =D/C$.  If we enforce the normalization Eq. (\ref{eq:17}),
we find that $D=B$.  This reduces the spatial probability
distribution to a function of two unknown parameters $B$
and $C$ assuming that $b$  is known.

We are at the stage where we can determine the number of annihilating
vortex pairs as a function of time.  
From Eqs.(\ref{eq:18}),  (\ref{eq:9}) and (\ref{eq:25})
we have that
\be
\gamma(t) =\frac{AB}{L^{1+b}}=\frac{AD}{L^{1+b}}
\label{eq:27}
\ee
and we again have that $D=B$.  However from Eq.(\ref{eq:22})
we have
\be
\frac{1}{1+b}\frac{d}{dt}L^{1+b}=AC
\ee
and for long times
\be
L^{1+b}=ACzt
~~~.
\ee
Putting this result back into Eq.(\ref{eq:27}) we find
\be
\gamma(t) =\frac{AB}{ACzt}=\frac{\alpha}{zt}
~~~.
\ee
We have then, from Eq. (\ref{eq:9}), that the number of pairs of vortices as a function of time is given by
\be
N(t)/N(t_{0})=\left(t_{0}/t\right)^{\alpha /z}
~~~.
\label{eq:31}
\ee
However from simple scaling ideas we have rather generally
that for a set of point defects in $d$ dimensions
\be
N(t)\approx L^{-d}
~~~.
\ee
Comparing this with Eq.(\ref{eq:31}) we identify $\alpha =d$.
This gives our final form for $F(x)$
\be \label{pr1}
F(x)=\frac{Bx^{b}}{(1+Cx^{1+b})^{1+d/z}}
~~~.
\ee
We check the validity of this result in sections III and IV.

The speed probability distribution is given by
\be
P_v(v,t)=\int_{0}^{\infty} dr~ P(r,v,t)
\nonumber
\ee
\be
=\int_{0}^{\infty} dr~\delta (v-Ar^{-b})P_r(r,t)
\nonumber
\ee
\be
=\frac{B}{AbL^{1+b}}\frac{1}{\tilde{v}^{2+1/b}}
\left(1+\frac{C}{L^{1+b}\tilde{v}^{(1+b)/b}}\right)^{-\sigma}
\ee
where $\tilde{v}=v/A$.  We can define the characteristic
speed $\bar{v}$ via
\be
\left(\frac{\bar{v}}{\tilde{v}}\right)^{(1+b)/b}
=\frac{C}{L^{1+b}\tilde{v}^{(1+b)/b}}
\ee
or
\be \label{av}
\bar{v}(t)=\frac{C^{(1+b)/b}}{L^{b}}\propto t^{-(1-1/z)}
\ee
and
\be
P_v(v,t)=\frac{B}{ACb\bar{v}}\frac{1}{V^{2+1/b}}
\left(1+V^{-(1+b)/b}\right)^{-\sigma}=\frac{1}{\bar{v}}\,p_v(v/\bar{v})
\ee
where $V=\tilde{v}/\bar{v}$ and the distribution function has a scaling form.  Clearly the large speed tail
goes as $V^{-p}$ where $p=2+1/b$ in agreement with
Bray's result.  After rearrangement we find
\be \label{pv}
P_v(v,t)=\frac{1}{\bar{v}}\,p_v(v/\bar{v})=\frac{d}{Ab\bar{v}}\frac{V^{s}}{(1+V^{(1+b)/b})^{\sigma}}
\ee
where 
\be
s=\frac{B}{Cb}-1=\frac{d}{b}-1
~~~.
\ee
We numerically test this result for various models below.

\section{Non-conserved n-vector model}

We now want to test our theoretical results for $P_r(r,t)$ and $P_v(v,t)$ for the non-conserved time-dependent Ginzburg-Landau (TDGL) $O(n)$ model where $b=z-1=1$. If we work in terms of dimensionless variables ${\widetilde v}=v/\bar{v}$ and ${\widetilde r}=r/\bar{r}$ where $\bar{v}(t)$ and $\bar{r}(t)$ are the average speed and separation as function of time, then  Eq. (\ref{pv}) gives the
vortex speed distribution function 
\begin{equation} \label{NCOPv}
p_v({\widetilde v}) =n\beta^{n/2}\frac{{\widetilde
v}^{n-1}}{\left(1+\beta\, {\widetilde
v}^2\right)^{(n+2)/2}}\ ,
\end{equation}
with $\beta=\pi[\Gamma(\frac{1+n}{2})/\Gamma(\frac{n}{2})]^2$. This is
exactly the familiar result found in Ref. \cite{Mazenko97} for $n=d$. The average speed
is $\bar{v}\propto t^{-1/2}$ and $z=b+1=2$.

As a special case, when $n=2$, we have 
\begin{equation} \label{ncn2}
p_v({\widetilde v}) = \frac{2\beta {\widetilde
v}}{\left(1+\beta\,{\widetilde v}^2\right)^2}\ ,
\end{equation}
with $\beta=(\pi/2)^2=2.4674$.  Both $p_v({\widetilde v})$ and $\bar{v}(t)$ have been
verified in Ref. \cite{Qian03}. The energy and
defect number are proportional to $(t/\ln t)^{-1}$, where there is a
logarithmic correction. But we did not see such a correction for the
average speed $\bar{v}(t) \propto t^{-1/2}$.

Let us turn next to the distance distribution function. From
Eq. (\ref{pr1}) we have, for $n=d=2$, 
\begin{equation} \label{pr2}
F({\widetilde r}) = \frac{2C{\widetilde r}}{(1+C{\widetilde r}^2)^2}\
,
\end{equation}
where $C=2.4674$.

We check this numerically using the same data as in Ref. \cite{Qian03}. The model is described by a
Langevin equation defined in a two-dimensional space
\begin{equation}
\frac{\partial \vec{\psi}}{\partial t} = \epsilon \vec{\psi} +\nabla^2
\vec{\psi} -(\vec{\psi})^2\vec{\psi}\ ,
\end{equation}
where $\epsilon $ is set to be
$0.1$, and the quench is to
zero temperature, so we need not include noise. We worked on $1024\times 1024$ system
with lattice spacing $\Delta r = \pi/4$. Periodic boundary conditions
are used. Starting from a completely disordered state, we used the Euler method to drive the system to evolve in time
with time step $\Delta t = 0.02$.

The position of a vortex is given by the center of its core
region, which is the set of points $(x_i,y_i)$ that satisfy $|\vec{\psi}(x_i,y_i)|<\langle |\vec{\psi}|\rangle/4$. By fitting $|\vec{\psi}(x_i,y_i)|$, where $(x_i,y_i)$ are the points belonging to a vortex's core region, to the function
$M(x,y)=A+B[(x-x_0)^2+(y-y_0)^2]$ we can find the center
$(x_0,y_0)$. The positions of  
each vortex at different times are recorded, and the speed is calculated 
using $v=\Delta d/\Delta \tau$. Here $\Delta d$ is the distance that the 
vortex travels in time $\Delta \tau= 5$. 

\begin{figure}[ht] 
\begin{center}\includegraphics[scale=0.31]{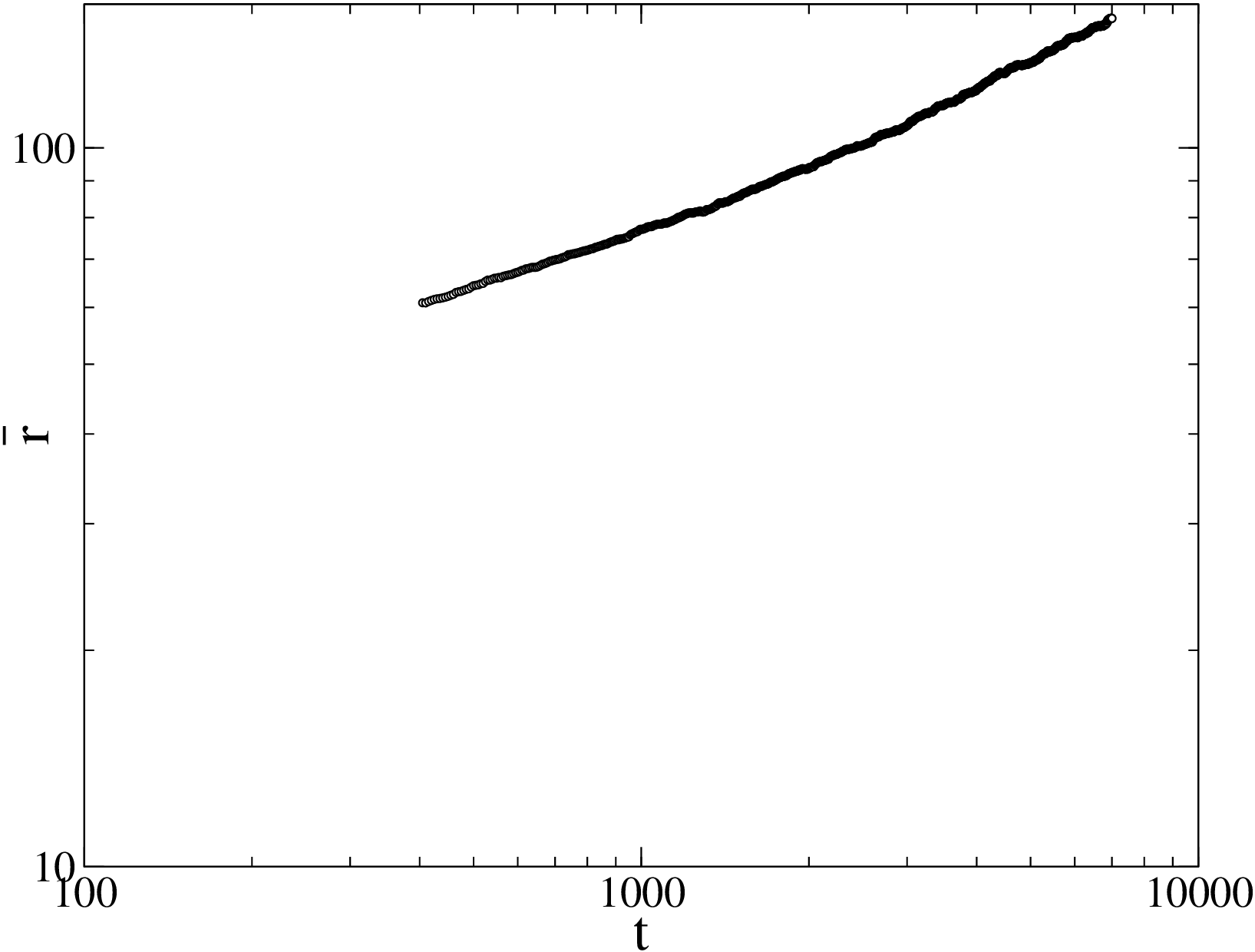}\end{center}
\caption{The average distance $\bar{r}$ between annihilating
pairs versus time after quench. The data is averaged over 68 runs.  The
fit  to the data is given by Eq. (\ref{avr1}).}
\label{avr}
\end{figure}

To measure $\bar{r}(t)$ and $F({\widetilde r})$ we must first accumulate
the following data. In a given run we keep track of the trajectories of
all the vortex centers. We label each pair of oppositely charged vortices which 
annihilate. Then move backwards in time to determine for each such pair
the separation as a function of time $r_i(t)$. Then $\bar{r}(t)$ is the
average separation between annihilating pair of vortices at time $t$. 
The average distance $\bar{r}(t)$ is shown in Fig. \ref{avr}.  From the discussion in Sec. II we expect
\begin{equation} \label{avr1}
\bar{r}(t)=L(t)=a\left[t+(\bar{r}_0/a)^z\right]^{1/z}\ ,
\end{equation}
where $\bar{r}_0=\bar{r}(t=0)$ and $a$ is a constant. The average
distance between annihilating pairs increases with time and a fit to the
data gives $a=2.71$, $\bar{r}_0\approx 50$ and $z\sim 2.22$.

To measure the probability distribution function $F({\widetilde r})$, we
distribute the various pairs into bins of width $\Delta = 0.01$ centered
about the scaled separation $r_i(t)/\bar{r}(t)$. We then plot the
number of pairs in each bin versus $r(t)/\bar{r}(t)$ and properly normalize to
obtain the scaling result shown in Fig. 2.  In the
following, when we measure the other distribution functions with scaling
properties we employ the same
method. 
 The curve
representing $F(x)$ given by  Eq. (\ref{pr2}) is also shown in Fig. 2. 
There is no free parameter in the fit 
other than $b$ and $\alpha$. The fit is fairly good. At large distances
we can see the function approximately obeys a power
law. The exponent is about $6$, which is different from the value 3
indicated by Eq. (\ref{pr2}). We do not know why it is so.

\begin{figure}[ht] 
\begin{center}\includegraphics[scale=0.31]{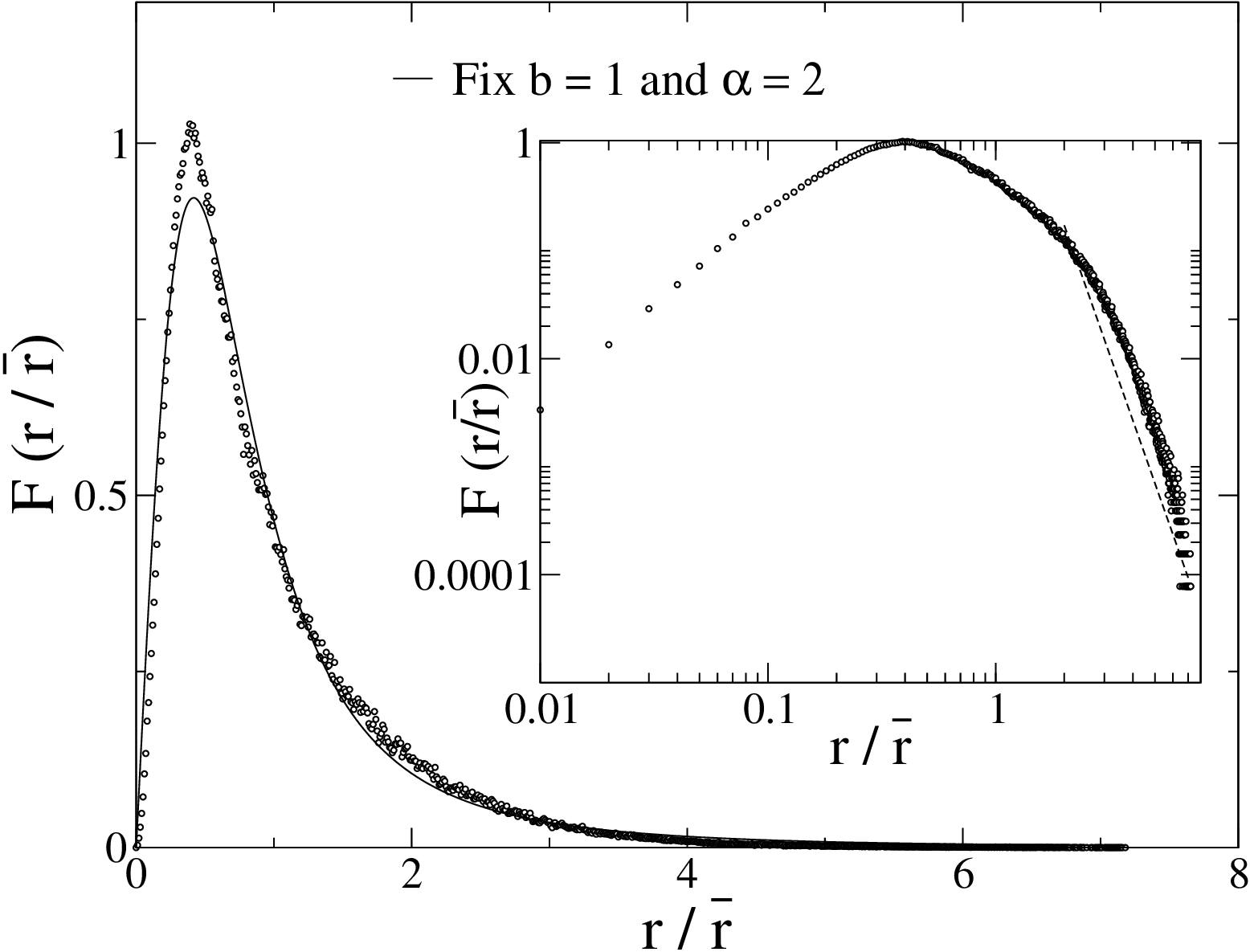}\end{center}
\caption{Separation probability distribution $F({\widetilde r})$ versus
the scaled separation for the
NCOP case with $n=d=2$.  The data is averaged over 68 runs with a bin size of 0.01. The solid line is
Eq. (\ref{pr2}) with $b=1$ and $\alpha=2$. In the insert we show the
same data on a logarithmic scale. At large $r$ the distribution is
approximately a power law with  an exponent about
$6$. The dashed line in the insert is
proportional to ${\widetilde r}^{-6}$.}
\label{fqr}
\end{figure} 

\begin{figure} 
\begin{center}\includegraphics[scale=0.31]{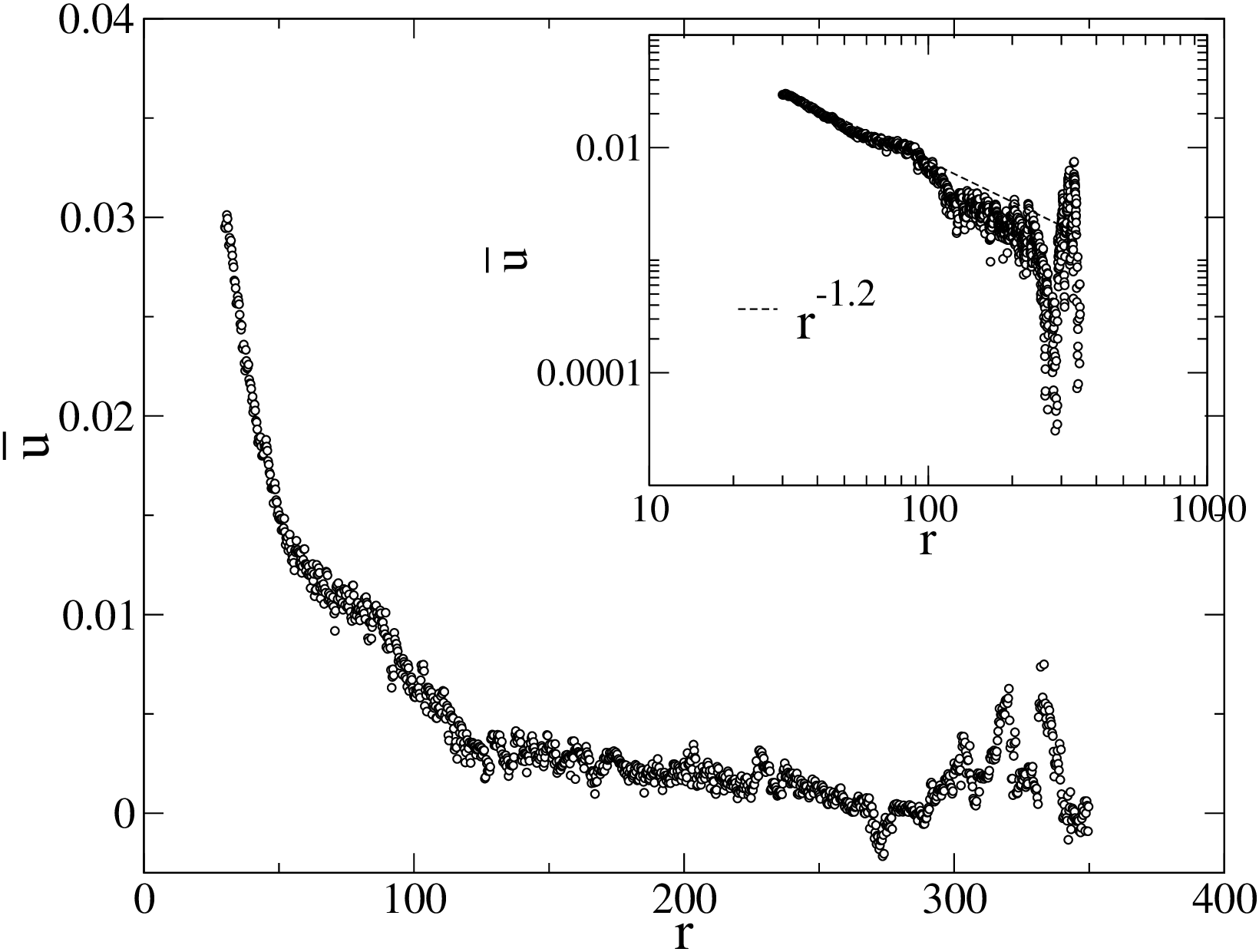}\end{center}
\caption{Test of Eq. (\ref{ur}). See text for a discussion. The data is averaged over 68 runs. 
There is a scaling regime at
small distances $r$ with an exponent near to $-1$. At large distances,
while the
statistics are not as good, there is still an approximately power-law dependence on $r$.}
\label{au}
\end{figure}

We next measure $\bar{u}(r)$, the average speed for annihilating defects
separating by a distance $r$. We track the motion of each annihilating
pair, and determine for each pair the speed $u_i=(r_i(t+\Delta
\tau)-r_i(t))/\Delta \tau$ as a function of $r_i$. Then we average
$u_i(r)$ over  all the pairs that have a fixed $r_i=r$. The result is
shown in Fig. \ref{au}. For small enough separations we have
$\bar{u}(r)\propto r^{-b}$ where $b\approx 1$ as expected.

\section{Conserved n-vector model}

Let us turn to the case of a conserved order parameter where for a TDGL
model we expect $b=z-1=3$. In this case the
vortex speed distribution, Eq. (\ref{pv}), is given by 
\begin{equation}
p_v({\widetilde v}) =\frac{n\beta^{n/4}}{3} \frac{{\widetilde
v}^{(n-3)/3}}{\left(1+\beta\, {\widetilde
v}^{4/3}\right)^{(n+4)/4}}
\end{equation}
with $\beta = \left(n\Gamma(\frac{5}{4})\Gamma(\frac{3+n}{4})/\Gamma(1+\frac{n}{4})\right)^{4/3}$and the average speed is $\bar{v}\sim t^{-3/4}$.
As a special case, consider $n=2$, where
\begin{equation} \label{cn2}
p_v({\widetilde v}) =
\frac{2\sqrt{\beta}}{3}\,\frac{{\widetilde v}^{-1/3}}{\left(1+\beta\,
{\widetilde v}^{4/3}\right)^{3/2}}\ , 
\end{equation}
with $\beta =2.27773...$. Notice, unlike the NCOP case, $p_v({\widetilde
v})$ blows up for small ${\widetilde v}$. This appears to be an
unphysical feature.

The distribution function for the distance between annihilating pairs,
Eq. (\ref{pr1}) with $b=3$ and $n=d=2$, gives
\begin{equation} \label{cpr2}
F({\widetilde r})=\frac{2C {\widetilde r}^3}{(1+C{\widetilde r}^4)^{3/2}}\ ,
\end{equation}
where $C=11.817...$. 

We simulated the conserved $O(2)$ model in two dimensions to test these predictions. The model is described
by a Langevin equation defined on a two-dimensional space
\begin{eqnarray} \label{dyneq}
\frac{\partial \vec{\psi}}{\partial t} &=& -\nabla^2 \vec{\psi} - \nabla^4
\vec{\psi} + \nabla^2 \left[ (\vec{\psi})^2\vec{\psi}\right]  \nonumber \\
&=&\nabla^2\frac{\delta {\mathcal H}_E[\vec{\psi}]}{\delta \vec{\psi}}\ ,
\end{eqnarray}
where the effective Hamiltonian is given by $\displaystyle{\mathcal H}_E[\vec{\psi}]=\int\left[-\frac{1}{2}\vec{\psi}^2+\frac{1}{2}(\nabla\vec{\psi})^2+\frac{1}{4}(\vec{\psi}^2)^2\right]d^2r$.  All the quantities
are dimensionless. We work on
$256\times 256$ system with lattice spacing $\Delta r = \pi/4$ and again
periodic boundary conditions are used. We employ the method invented by
Vollmayr-Lee and Rutenberg \cite{VR03} to numerically integrate
Eq. (\ref{dyneq}). This method is stable for any value of integration
time step $\Delta t$. As the time $t$ increases, the evolution of
the system becomes progressively slower. With the new time step technique we can increase
the time step to accelerate the evolution. We let $\Delta t =
0.01\,t^{0.36}$ after $t>120$. 

\begin{figure}[ht] 
\begin{center}
\includegraphics[scale=0.31]{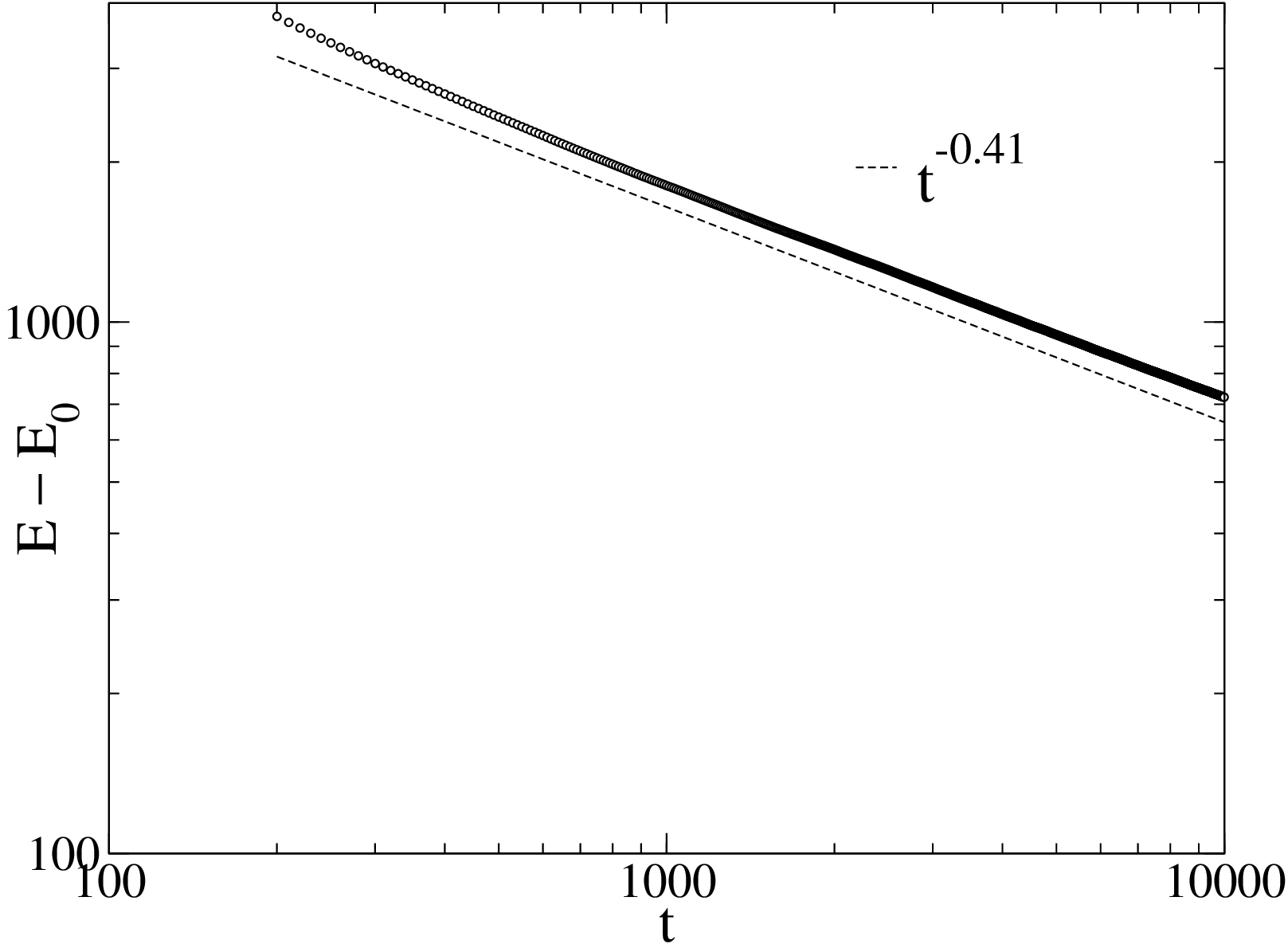}

\vspace{1mm}

\includegraphics[scale=0.31]{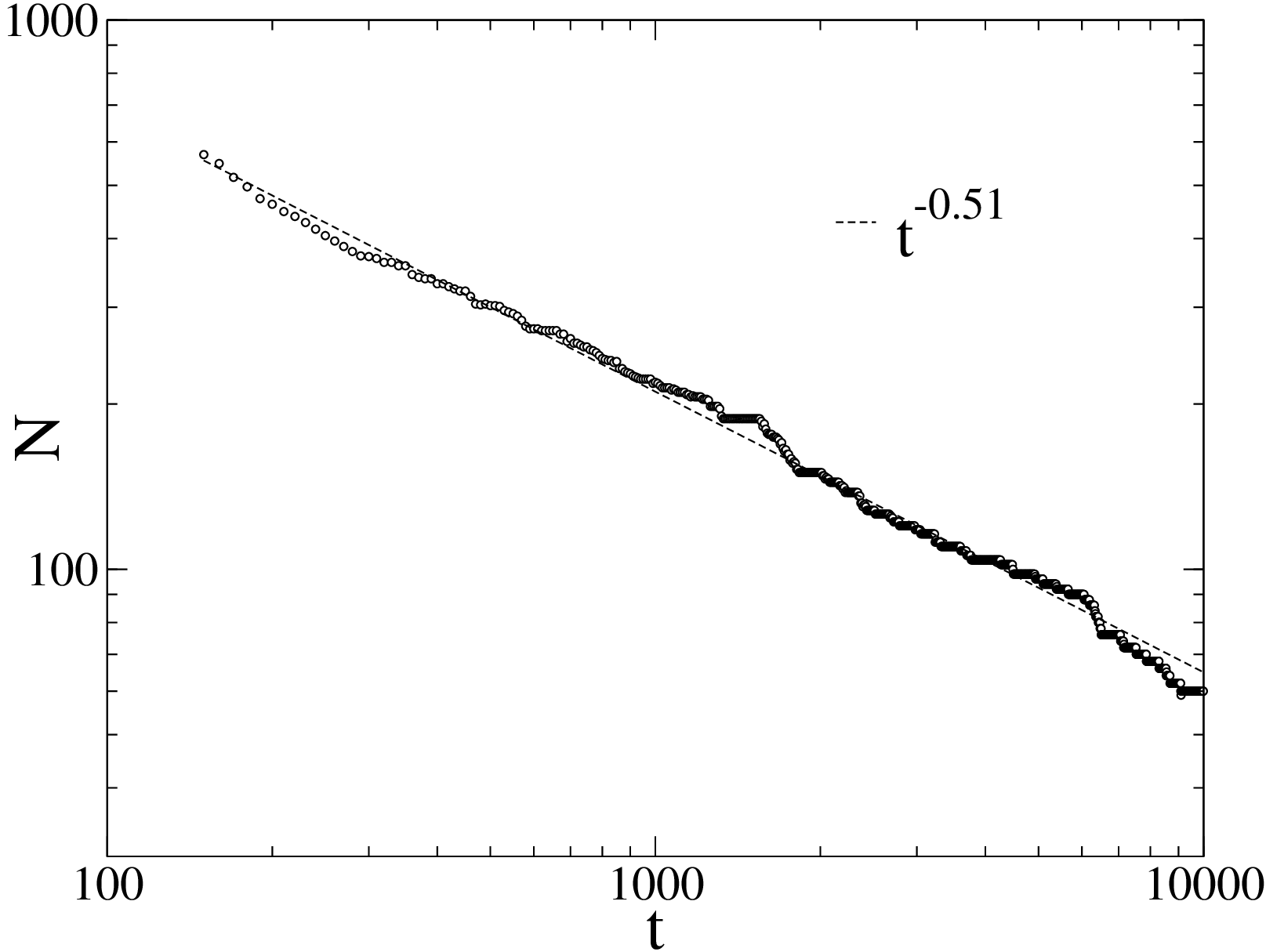}
\end{center}
\caption{The energy $E(t)-E_0$ and the vortex number $N(t)$ are plotted
versus time after quench. The data is
averaged
over 61 runs.  The dashed lines, which are guides to the eye, are
proportional to $t^{-0.41}$ for the energy plot and $t^{-0.51}$ for the
vortex number plot respectively.}
\label{eng}
\end{figure}

In addition to the vortex statistics discussed in the NCOP case, we also measure the average energy $E=\langle{\mathcal H}_E\rangle$  above the ground state energy $E_0 = -S/4$ with $S$ being the area of the system.
The energy and number of vortices are shown in Fig. \ref{eng}. The
power-law exponent for the defect number is $-0.51$, which is consistent
with $d/z=1/2$. The decay power law for the energy is $t^{-0.41}$, which
is slower than that for the defect number. This may be due to the
relaxation of spin waves.

\begin{figure}[ht] 
\begin{center}\includegraphics[scale=0.31]{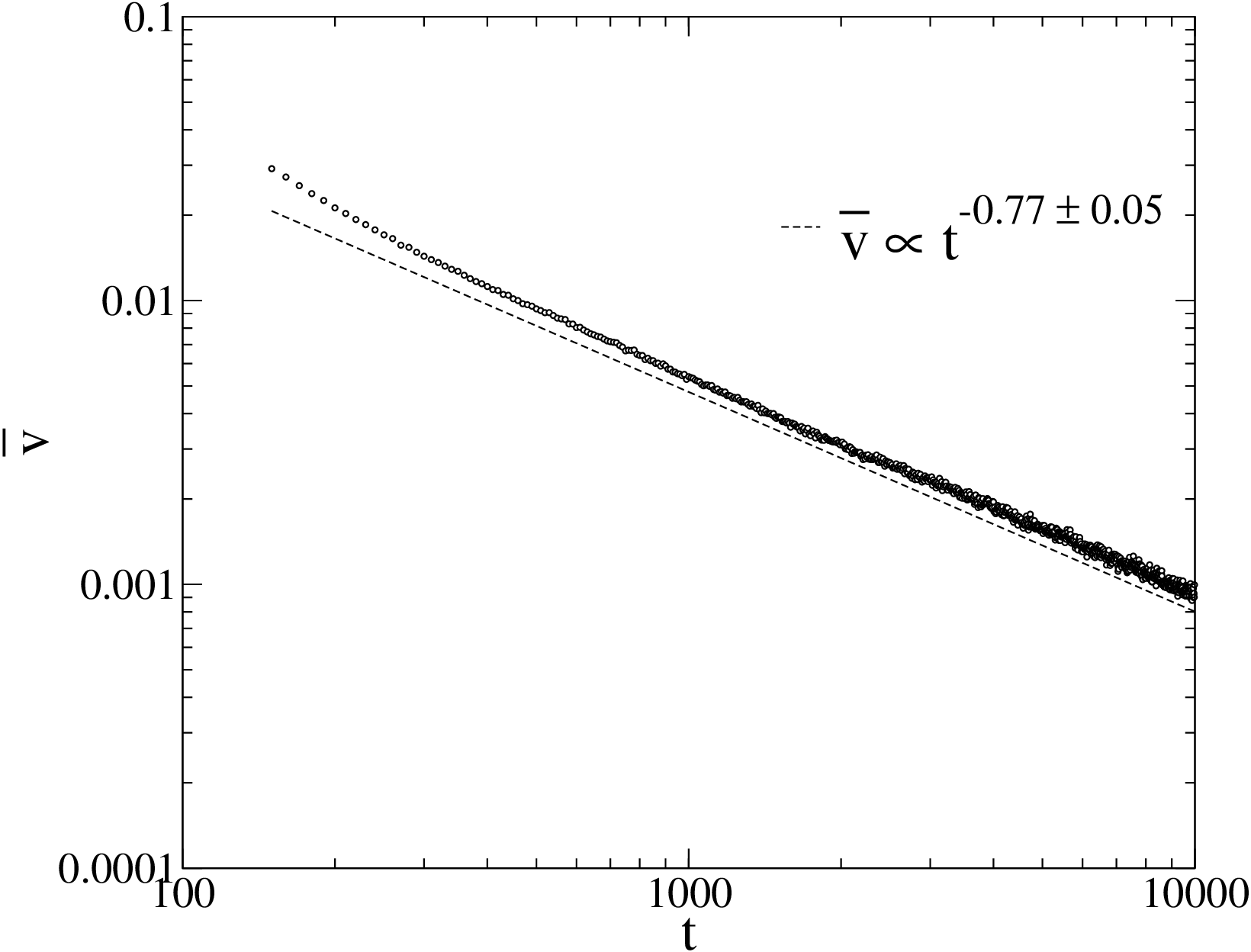}\end{center}
\caption{The averaged speed of vortices at time $t$ for the COP
case. The data is averaged over 61
runs. }
\label{cav}
\end{figure}

We use the same method as in the non-conserved case to find the
center for each vortex. The speed of each vortex is computed by using $v=\Delta
d/\Delta \tau$ with $\tau = 10$. 
 We measure the speed for each vortex at the same time $t$ and average
over different vortices. The average speed of the defects is shown in
Fig. \ref{cav}. The
prediction for the exponent is $-(1-1/z)=-0.75$, while the measurement finds
$-0.77$. 

\begin{figure}[ht] 
\vspace{3mm}
\begin{center}\includegraphics[scale=0.3]{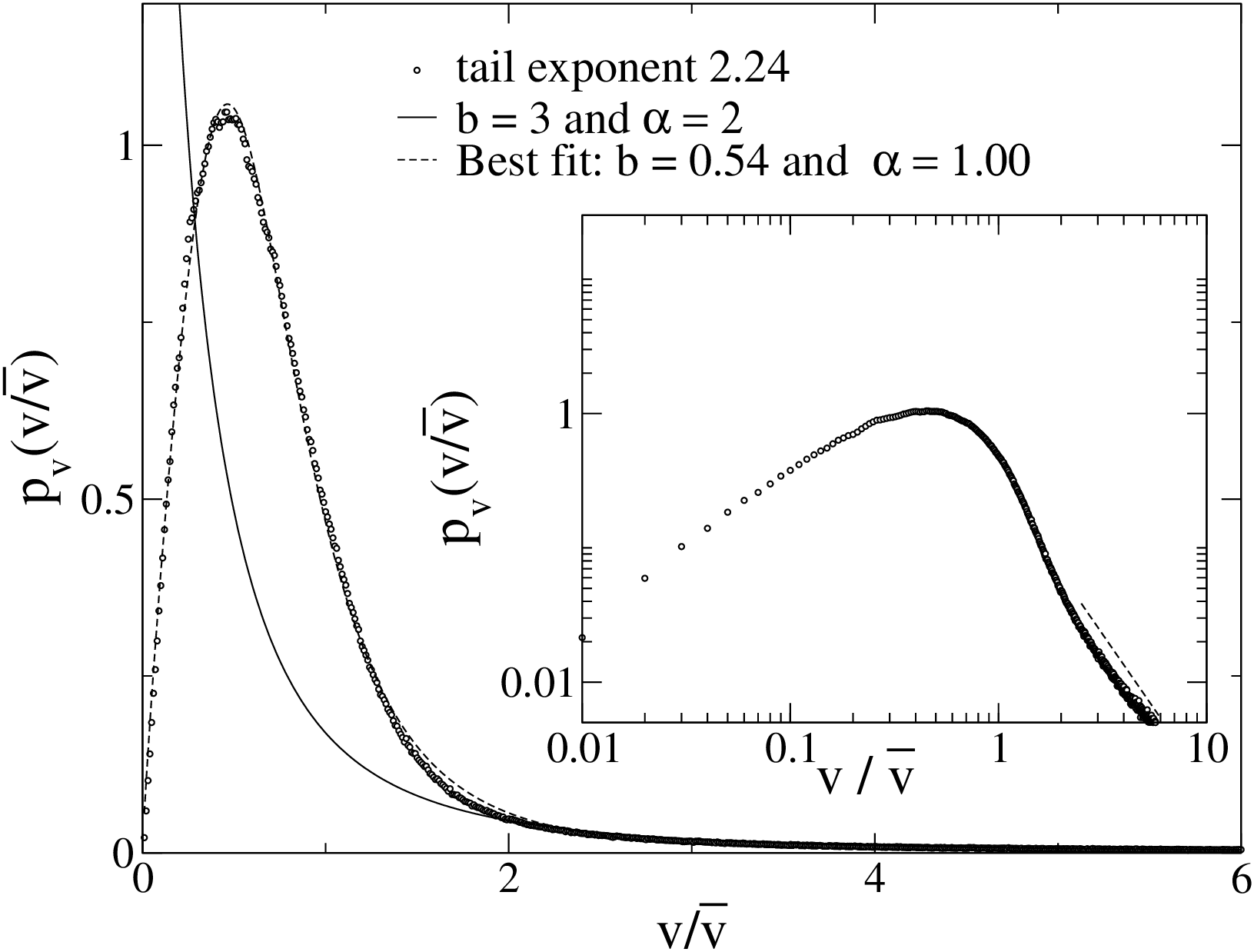}\end{center}
\caption{The probability distribution for vortex speed for the COP case. The data are
from 61 different runs. The bin size is 0.01. The
solid  line is Eq. (\ref{cn2}). The dashed line is the best fit to
$p_v({\widetilde v})$ by changing $b$ and $\alpha$. In the insert we show
the same data on a logarithmic scale. The dashed line in the insert is used to guide the
eye and proportional to  ${\widetilde v}^{-2.24}$. }
\label{cfq}
\end{figure}

Next we determine the speed distribution as a function of time. Again we
plot the scaled data from
different times to test the scaling property of this distribution
function. The resultant $p_v({\widetilde v})$ is shown in
Fig. \ref{cfq}. Clearly scaling works and the large speed tail exponent
is $2.24$. This is close to the prediction $2+1/b=2.33$. However the
theory fails at small scaled speeds where the simulations go to
zero while the theory blows up. Clearly the exponent $s$ in Eq. (38) is
poorly determined in the theory for $b=2$ and $d=2$. If we allow $b$ and
$\alpha$ float then we obtain an excellent fit shown in Fig. 6.

\begin{figure}[ht] 
\begin{center}\includegraphics[scale=0.31]{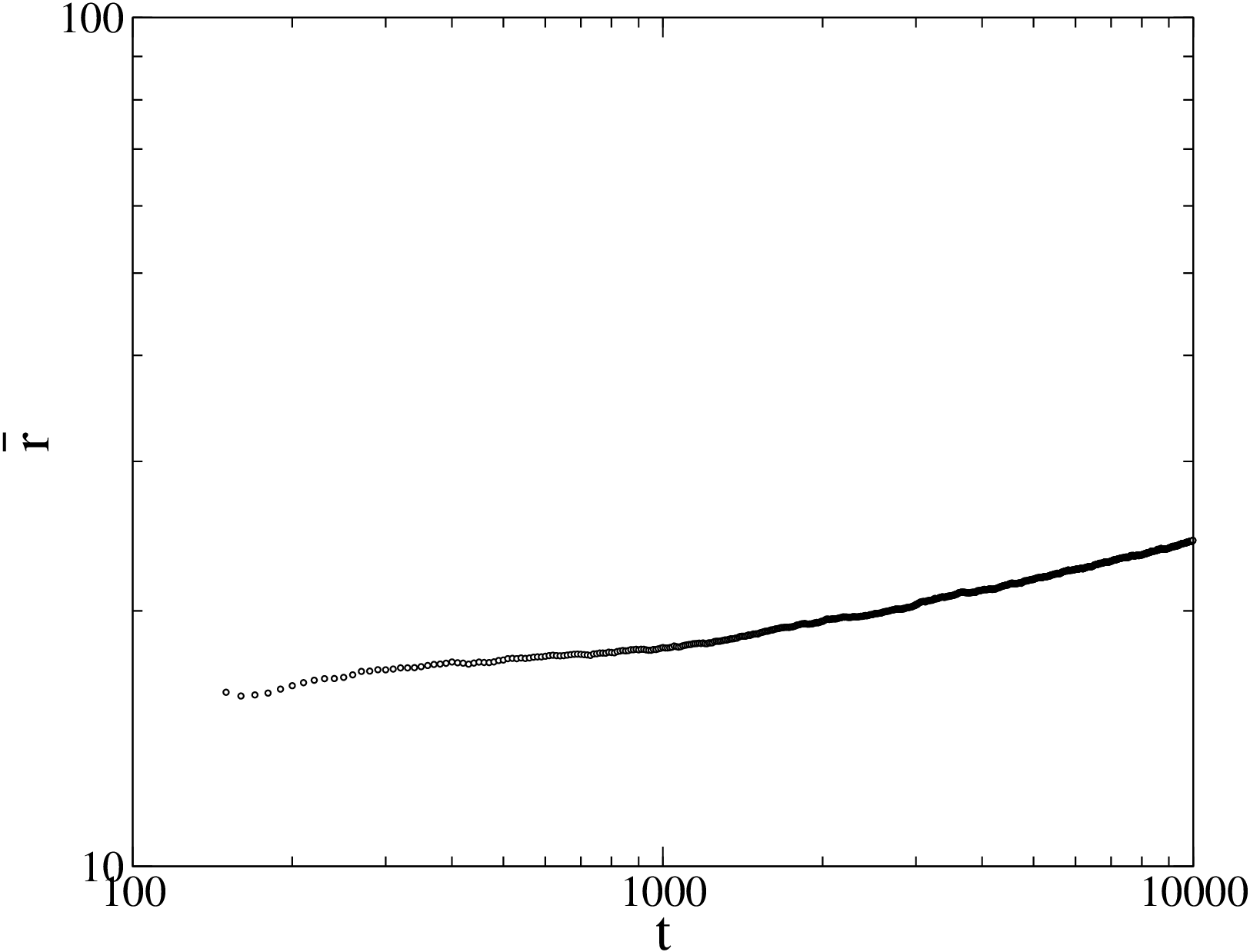}\end{center} 
\caption{The average distance $\bar{r}$ between annihilating
pair of vortices versus time after quench for the COP case. The data is
from 61 runs. The fit to Eq. (\ref{avr1}) is given in the text.}
\label{cavr}
\end{figure}

The average separation of annihilating pairs of vortices $\bar{r}(t)$
for the COP case is 
shown in Fig. \ref{cavr}. Again fitting this to the form given by
Eq. (\ref{avr1}) we find $a=2.285$, $\bar{r}_0 = 17$ and $z=4.0$.

\begin{figure}[ht] 
\begin{center}\includegraphics[scale=0.31]{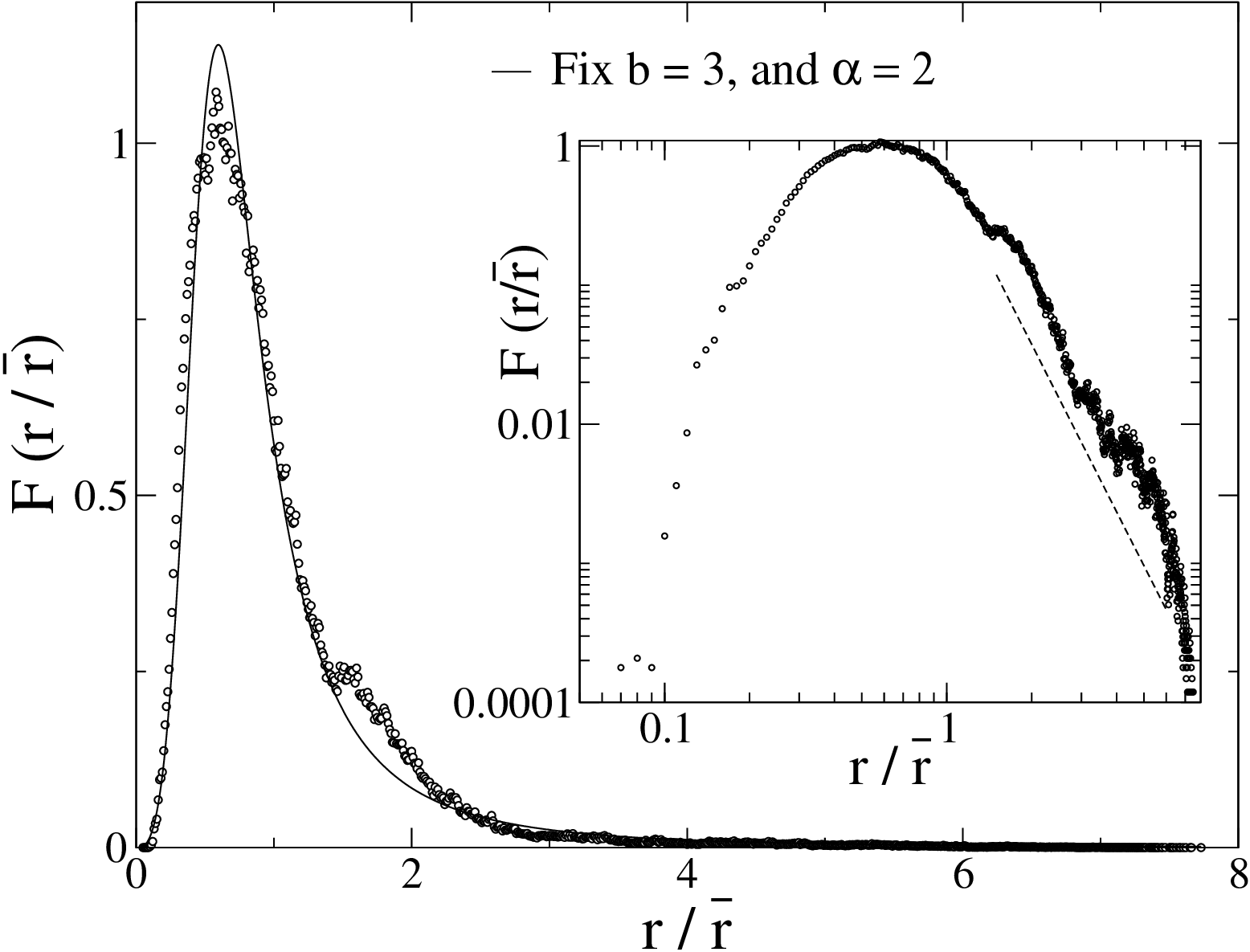}\end{center}
\caption{The probability distribution for the separation between two
annihilating vortices for the COP case. The data is averaged over 61 runs. Bin size is 0.01. The solid line is
Eq. (\ref{cpr2}) with $b=3$ and $\alpha = 2$. In the insert we show the
same data with a logarithmic scale. The behavior of this function at
large distances is approximately a power law with an exponent about
$4$. The dashed line is proportional to
${\widetilde r}^{-4}$.} 
\label{cfqr}
\end{figure}

As in the NCOP case we can measure the separation distribution function $F({\widetilde r})$. This is shown in
Fig. \ref{cfqr}. With no free parameters, $b$ and $\alpha$ being
fixed, the fit is pretty good. At large distances, the statistics are
poor.  But we can see the function approximately obeys a power
law. The exponent is about $4$, which is different from the value 3
indicated by Eq. (\ref{cpr2}). 

\begin{figure}[ht] 
\begin{center}\includegraphics[scale=0.31]{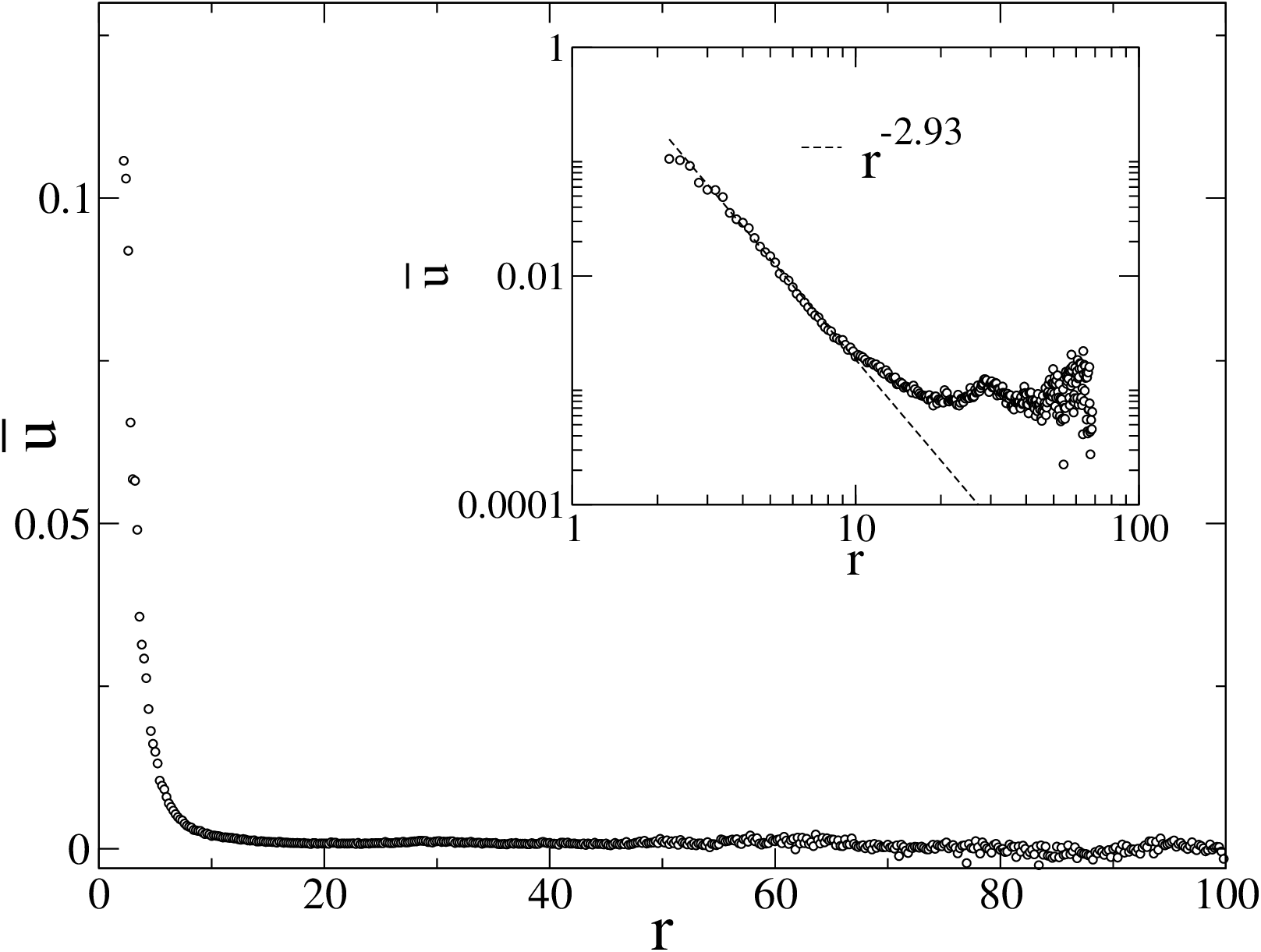}\end{center}
\caption{The average speed as a function of pair separation $r$ for the COP case. There is a scaling regime at
small $r$ with the exponent near to $b=3$. The data is collected over 61 runs. }
\label{cau}
\end{figure}

We also measure $\bar{u}(r)$, as in the NCOP case. Our results are shown in Fig. \ref{cau}. The assumption $u\sim r^{-b}$ is well satisfied with $b\approx 3$ at small enough distances.

\section{Scalar models}

We follow the argument given by Bray \cite{Bray97} to extend our
discussion to include models
with a scalar order parameter ($n=1$). First we calculate the
probability function $P_r(r,t)$ for the domains with radius $r$. This
calculation is the same with that in Sec. II. We obtain Eq. (33) for
$F(x)$. Next, we compute the area-weighted probability for interfacial
radius of curvature $r$ by multiplying $F(x)$ with $x^{d-1}$ and then
normalize the resulting quantity. The resulting probability function $F_s(x)$ is
\begin{equation}
F_s(x) = \frac{d\Gamma(d/z)C^{(1+d+z)/z}}{\Gamma(1/2)\Gamma((d+z-1)/z)} \frac{x^{d+z-2}}{(1+C x^z)^{1+d/z}}\ .
\end{equation}
Following Bray we use $v=Ar^b$ to get the distribution function for the interface speed
\begin{equation}
P_v({\widetilde v}) = \frac{d\Gamma(d/z)\beta^{1/z}}{(z-1)\Gamma(1/z)\Gamma((d+z-1)/z)}\frac{{\widetilde v}^{-1+1/(z-1)}}{(1+\beta {\widetilde v}^{1+1/(z-1)})^{1+d/z}}\ .
\end{equation}
For the non-conserved case, this result is the same with the one
obtained by Bray using the Gaussian calculation. The large speed
tail  exponent is $p=2+d/(z-1)$, which is valid for both conserved and non-conserved models. 

\section{Conclusion}

We show how a simple generalization  of  Bray's scaling argument can
lead to quantitative results for certain distribution functions. In
particular  we find that the distribution function for the distance
between  annihilating pairs
of vortices is well described by the scaling theory for both NCOP and
COP  dynamics for $n=d=2$. We are also able to
compute the speed distribution function using these ideas.  For
non-conserved models, we reproduce the accurate result obtained previously. For conserved
models, the speed distribution function only gives us the correct tail
exponent. 

Our method can also be extended to scalar cases, and generate a full
expression for the interfacial speed distribution. The power-law tail
exponent $p=2+d/(z-1)$ is obtained. The result is the same as the result obtained by Bray \cite{Bray97}.  

The simple scaling method presented here leads to a reasonable
description of the statistics of defect dynamics. Clearly it does a
better job for the NCOP case since the speed distribution function for
COP case does not show the proper small speed behavior. Similarly the
more microscopic method of Ref. \cite{Mazenko03} leads to an adequate
treatment of the small speed regime for the COP case but does not give
the correct large speed tail.

The approach developed here is highly heuristic. Can it be systematized?     
Clearly to improve this approach one would need to include the
interactions between different pairs. It is not clear how one does
this. 

Acknowledgement: This work was supported by the National Science
Foundation under contract No. DMR-0099324 and by the Materials Research
Science and Engineering Center through Grant No. NSF DMR-9808595.


\begin{thebibliography}{99}
\bibitem{foot} More generally Eq. (\ref{eq:6}) has the solution:
\begin{equation}
P_r(r,t)=\frac{1}{u(r)}\exp\left(\int_{t_*}^t\gamma(t')\,dt'\right)\,\chi(t-\tau(r))\
, \nonumber
\end{equation}
where $\displaystyle\tau(r)=\int_{r_*}^r\frac{dr'}{u(r')}$ and $\chi$ is an arbitrary function, $r_*$ and $t_*$ are two constants.
\end{thebibliography}
\end{document}